\pdfoutput=1
\documentclass{article}

 \usepackage[final]{neurips_2019_ml4ps}

\usepackage[utf8]{inputenc} %
\usepackage[T1]{fontenc}    %
\usepackage{hyperref}       %
\usepackage{url}            %
\usepackage{booktabs}       %
\usepackage{amsfonts}       %
\usepackage{nicefrac}       %
\usepackage{microtype}      %
\usepackage{graphicx}
\usepackage{subcaption}
\usepackage{tikz}
\usetikzlibrary{shapes.geometric, arrows}
\usepackage{float}
\usepackage{caption}
\usepackage[labelformat=simple]{subcaption}
\usepackage{sidecap}
\usepackage{graphicx}
\usepackage[export]{adjustbox}
\usepackage{amssymb}

\setcitestyle{square,aysep={},yysep={;}}

\title{3D Conditional Generative Adversarial Networks to enable large-scale seismic image enhancement}

\author{
  Praneet Dutta$^{1}$, Bruce Power$^{2}$, Adam Halpert$^{2}$, Carlos Ezequiel$^{1}$, Aravind Subramanian$^{3}$ \\
  \textbf{Chanchal Chatterjee$^{1}$,  Sindhu Hari$^{3}$,
  Kenton Prindle$^{1}$, Vishal Vaddina$^{3}$,  Andrew Leach$^{1}$}\\
  \textbf{  Raj Domala$^{3}$, Laura Bandura$^{2}$, Massimo Mascaro$^{1}$}\\
  $^{1}$Google AI\\
  $^{2}$Chevron Energy Technology Company\\
  $^{3}$Quantiphi\\
  \texttt{\{praneetdutta, cezequiel, cchatterjee,kentonprindle andrewleach, massy\}@google.com}\\
  \texttt{\{brucepower, halpert, laura.bandura\}@chevron.com}\\
  \texttt{\{aravind.subramanian, sindhu.hari, vishal.vaddina, raj.domala\}@quantiphi.com}\\
}
\sidecaptionvpos{figure}{c}

\begin{document}

\maketitle

\begin{abstract}

We propose GAN-based image enhancement models for frequency enhancement of 2D and 3D seismic images. Seismic imagery is used to understand and characterize the Earth's subsurface for energy exploration. Because these images often suffer from resolution limitations and noise contamination, our proposed method performs large-scale seismic volume frequency enhancement and denoising. The enhanced images reduce uncertainty and improve decisions about issues, such as optimal well placement, that often rely on low signal-to-noise ratio (SNR) seismic volumes. We explored the impact of adding lithology class information to the models, resulting in improved performance on PSNR and SSIM metrics over a baseline model with no conditional information.

\end{abstract}

\section{Introduction}

In geophysical imaging, resolution limitations of the seismic migration methods are well-known~\citep{Beylkin1985, doi:10.1190/1.1444602}. At depths typical of modern exploration targets, the seismic wavelength can be in excess of 250 meters, meaning that geo-scientists may be unable to resolve individual rock formations less than 60 meters thick, which is needed to be successful in exploration settings. In addition, variations in lithology and features such as faults can cause further disruption and attenuation of the acoustic wave energy. As a result, interpreting the underlying geological model from these image volumes has high uncertainty and has been the focus of much of the current and ongoing energy resource exploration. These limitations make purely data-driven approaches such as image super-resolution techniques to enhance seismic images more attractive.

Image super-resolution is the process of converting low-resolution images into high-resolution ones. The use of deep neural network approaches to image super-resolution is a growing research area with real-world applications in various fields~\citep{DBLP:journals/corr/abs-1902-06068}. Promising attempts at image super-resolution through deep neural networks include use of a pixel-recursive method~\citep{DBLP:journals/corr/DahlNS17} and zero-shot learning~\citep{DBLP:journals/corr/abs-1712-06087}. A survey of approaches can be found in ~\citep{DBLP:journals/corr/abs-1902-06068}. Early attempts to use deep learning and GANs for seismic image enhancement have also shown promise~\citep{doi:10.1190/segam2018-2996943.1}. 

Our proposed seismic image enhancement approach expounds on the work of~\citep{DBLP:journals/corr/LedigTHCATTWS16} that describes a super-resolution generative adversarial network (SRGAN) model, which we adapt to support both 2D slices of a 3D seismic cube and 3D cube partitions. Additionally, we employ pixel-level class conditional information based on geological lithology to improve our performance.

\

\section{Image enhancement approach}\label{Approach}
We add conditional information in the form of additional image channels that determine the lithology class associated with each pixel. We considered two methods in representing lithology class information: deterministic and probabilistic. In the deterministic method, the classes are represented as a one-hot encoded vector for each corresponding image pixel with a total of thirty one possible classes. In the probabilistic method, the class channel represents the binary probability of a pixel belonging to the salt class.

\begin{figure}[H]
    \centering
    \includegraphics[width=1\linewidth]{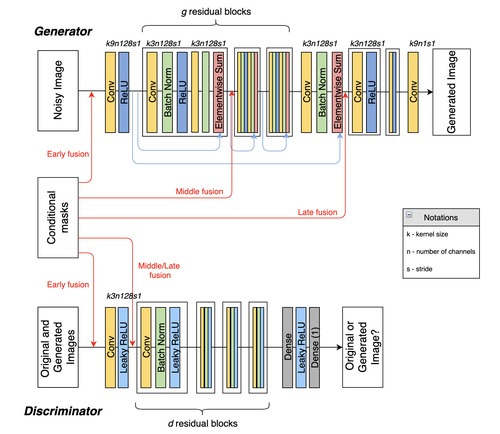}
    \caption{SRGAN-based model architecture for enhancing seismic images. In this work, the depth of the generator and discriminator is represented as a function of the number of repeating residual blocks.}
    \label{fig:ModelArch}
\end{figure}

To incorporate the class conditional information into SRGAN, we considered multiple fusion locations, such as early, mid, and late fusion~\citep{Snoek:2005:EVL:1101149.1101236} and fusion types, such as concatenation and dot product of the conditional information with ground truth (depicted in Figure~\ref{fig:ModelArch}). Early fusion meant augmenting the class conditional information at the generator/discriminator input layer. Mid fusion meant adding the information at the first residual block of the generator, and before the first residual block of the discriminator. Late fusion happened after all the repeating residual blocks of the generator, and in the same place as mid fusion for the discriminator. Since input and output images needed to have the same dimensions, we omitted an upsampling layer in the generator.

\subsection{Architecture}

\subsection{Loss function}

We base our loss function on the work of~\citep{DBLP:journals/corr/LedigTHCATTWS16} that defines a perceptual loss function as a weighted sum of content loss (based on MSE loss) and adversarial loss respectively.

\begin{equation}\label{eq:ContentLoss}
    l_{MSE} = \frac{1}{w \cdot h}\sum_{i=1}^{i=w}\sum_{j=1}^{j=h}(x_{i,j}-G(z|c)_{i,j}))^{2}
\end{equation}

In Equation~\ref{eq:ContentLoss}, $w$ and $h$ refer to image pixel width and height respectively, $G$ is the generator function that takes as input both the noisy image $z$ and conditional information $c$, and $x$ is the ground truth image. Minimizing MSE loss maximizes peak signal-to-noise ratio (PSNR), which is a commonly used image quality estimation. The adversarial loss is based on~\citep{1411.1784} that adds extra conditional information to the two-player minimax game with value function $V(D, G)$ originally proposed in~\citep{1406.2661}.

\begin{equation}\label{eq:AdversarialLoss}
 \min_{G}\max_{D}V(D, G) =  \mathbb{E}_{p} [log(D(x|c)) + log(1 - D(G(z|c)))]
\end{equation}

where $p$ is the joint distribution of $x$, $y$, and $c$. Note that this differs from standard GAN formulations in that the noise is not explicitly sampled, but is a result of the observation process of the ground truth $x$ that yields $z$ and $c$.
$D$ refers to the discriminator function, which pushes the generator to output images in the enhanced seismic image manifold. We further direct the generation process of $G$ by conditioning both $G$ and $D$ with additional information $c$ that refers to the lithology classes in our study.

\section{Experiments}\label{Experiments}

\subsection{Seismic Dataset}\label{Dataset}

\begin{figure}
    \includegraphics[width=0.66\linewidth]{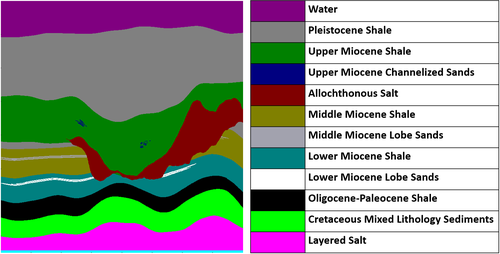}
    \centering
    \caption{Geological lithology classes for conditioning GAN prediction.} 
    \label{fig:GeoClasses}
\end{figure}

More than 100 000 training images were extracted from the SEAM I dataset~\citep{doi:10.1190/1.9781560802945}. This seismic data is a result of a finite difference forward model where a simulation of an acoustic wave field is propagated through the earth model volume computationally. The original earth model is also a synthetic model that was designed to effectively reproduce the actual lithology and structure found in the earth’s subsurface in sedimentary basins where energy exploration and production occurs.  As a result, the lithologies and labels of the rock properties are known down to the pixel scale. The lithology classes considered for the study are shown in Figure~\ref{fig:GeoClasses}. 

To generate the degraded seismic image input, we used a 5Hz low-pass filter and added 50\% uniform random noise.

\begin{figure}[h!]
  \centering
  \captionsetup{justification=centering}
  \begin{subfigure}[t]{0.33\linewidth}
    \includegraphics[width=\linewidth,height=2in]{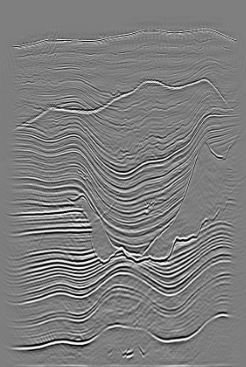}
    \caption{Ground truth seismic data (full bandwidth).}
    \label{fig:GroundTruth}
  \end{subfigure}
  \begin{subfigure}[t]{0.33\linewidth}
    \includegraphics[width=\linewidth,height=2in]{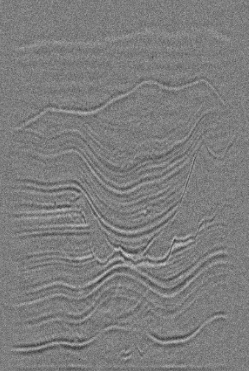}
    \caption{Degraded input (filtered and additive noise).}
    \label{fig:Degraded}
  \end{subfigure}
  \begin{subfigure}[t]{0.33\linewidth}
    \includegraphics[width=\linewidth,height=2in]{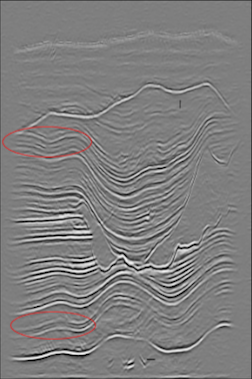}
    \caption{Generated output with no conditional information.}
    \label{fig:GenNoConditional}
  \end{subfigure}
  \begin{subfigure}[t]{0.33\linewidth}
    \includegraphics[width=\linewidth,height=2in]{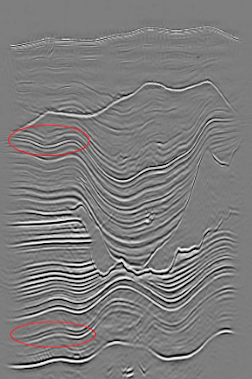}
    \caption{Generated output with conditional information.}
    \label{fig:GenConditional}
  \end{subfigure}
  \caption{Comparison of generated seismic images (bottom) with ground truth \subref{fig:GroundTruth} and degraded input \subref{fig:Degraded} (top). Red ovals in \subref{fig:GenNoConditional} and \subref{fig:GenConditional} highlight two regions of improved data resolution from using geological conditioning information.}
  \label{fig:Comparison}
\end{figure}

\begin{figure}[h!]
    \centering
    \captionsetup{justification=centering}
    \begin{subfigure}[t]{0.28\linewidth}
    \includegraphics[width=\linewidth, height=5cm,valign=t]{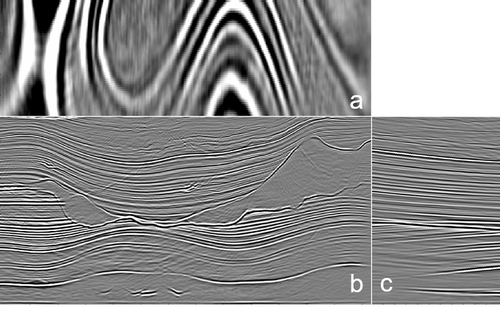}
    \caption{3D image ground truth.}
    \label{fig:CrossGroundTruth}
   \end{subfigure}
   \begin{subfigure}[t]{0.28\linewidth}
    \includegraphics[width=\linewidth, height=5cm, valign=t]{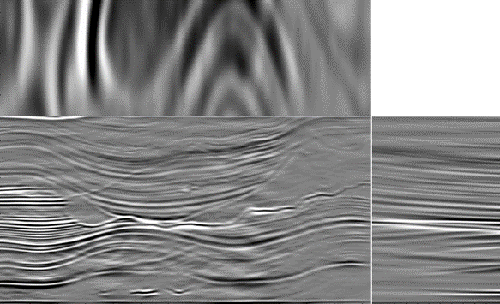}
    \caption{Degraded 3D image used as input data.}
    \label{fig:CrossInput}
   \end{subfigure}
  \begin{subfigure}[t]{0.28\linewidth}
    \includegraphics[width=\linewidth, height=5cm, valign=t]{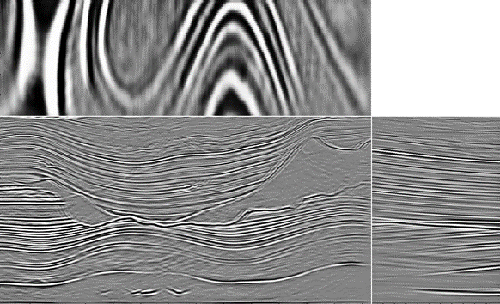}
    \caption{Enhanced 3D image.}
    \label{fig:CrossModel}
 \end{subfigure}
 \caption{Cross-sections of an image volume for ground truth~\subref{fig:CrossGroundTruth}, degraded input~\subref{fig:CrossInput} and model output~\subref{fig:CrossModel} respectively. The image in~\subref{fig:CrossGroundTruth} identifies the three different transects through the cube - plane view \textit{a}, x-y cross-section \textit{b}, and z-y cross-section \textit{c}.}
 \label{fig:Comparison3D}
\end{figure}

\subsection{Metrics}\label{Metrics}

To validate the model, we considered the following objective image quality metrics: peak signal-to-noise ratio (PSNR), structural similarity index (SSIM) and multi-scale SSIM (MS-SSIM). The multi-scale method (MS-SSIM) provides more flexibility in that it can incorporate image details at different resolutions.

\subsection{Results for 2D}

The outputs of our 2D models were visually examined by domain experts, who verified the efficacy of the approach (see Figure~\ref{fig:Comparison}). The results were visually examined to determine if reflection amplitude, phase, and coherence were consistent with the high frequency image as well as the underlying earth model.
Table~\ref{tbl:2DBest} summarizes the best results we obtained from the different 2D image enhancement models. We achieved the best result using a model trained with probabilistic conditional information (second row of Table~\ref{tbl:2DBest}). In our experiments, conditional models appear to have performed better than Baseline SRGAN  in most cases, regardless of fusion strategy.

\begin{table}[h!]
    \centering
    \caption{Summary of best results for 2D}
    \label{tbl:2DBest}
    \resizebox{\textwidth}{!}{%
    \begin{tabular}{lllllllll}
        \hline
        Model                  & Generator Depth & Fusion Type & Fusion Pos & MS-SSIM & SSIM & \%SSIM Gain & PSNR & \% PSNR Gain \\ \hline
        Baseline SRGAN          & 16              & -              & -       & -      & 0.549          & - & 20.47      & - \\
        \textbf{Probabilistic} & \textbf{32}     & \textbf{Concat} & \textbf{Early} & \textbf{0.784} & \textbf{0.656} & \textbf{19.48} & \textbf{22.97} & \textbf{12.21} \\
        \textbf{Deterministic} & \textbf{32}     & \textbf{Dot}   & \textbf{Mid}    & \textbf{0.785} & \textbf{0.642} & \textbf{16.93} & \textbf{22.96} & \textbf{12.16} \\
        Deterministic & 8     & Dot & Late & 0.760 & 0.600 & 09.28 & 22.13 & 08.11 \\
        Probabilistic & 8     & Concat & Early & 0.741 & 0.575 & 04.73 & 20.87 & 01.95 \\\hline
    \end{tabular}%
    }
\end{table}

\subsection{Results for 3D}

\begin{table}[h!]
    \caption{Summary of results for 3D}
    \label{tbl:3D}
    \centering
    \resizebox{\textwidth}{!}{%
    \begin{tabular}{@{}llllllll@{}}
        \hline
        Model & Fusion Type & Fusion Pos & SSIM & SSIM \% Gain & PSNR & \% PSNR Gain \\ \hline
        Baseline SRGAN & - & - & 0.94 & - & 29.95 & - \\
        \textbf{Deterministic} & \textbf{Concat} & \textbf{Late} & \textbf{0.98} & \textbf{4.25} & \textbf{33.22} & \textbf{10.91} \\
        Deterministic & Concat & Early & 0.95 & 1.06 & 30.40 & 1.5 \\ \hline
    \end{tabular}%
}
\end{table}

Table~\ref{tbl:3D} summarizes the results of the different 3D image enhancement models, with the best result obtained using a model trained with deterministic information and late fusion. Even without conditional information in the Baseline SRGAN model, we were still able to achieve an SSIM of 0.94. Figure~\ref{fig:Comparison3D} provides a visual comparison of the 3D model outputs by showing orthogonal cross-sections through the 3D image volume. For models trained with deterministic information, late fusion with concatenation provided the best results.

\section{Conclusion and Future Work}\label{Conclusion}
We were able to achieve improved performance on seismic image enhancement tasks using an SRGAN-based model with conditional information over one that did not have such information for 2D and 3D cases. Models trained with probabilistic information performed better than models trained on deterministic information on all metrics used in this study. In spite of this, the Baseline SRGAN model was still able to produce enhanced seismic images that were visually similar to ground truth.

\bibliographystyle{plainnat}
\bibliography{camera_ready}

\section{Appendix} 
\subsection{Acknowledgements}
We would like to acknowledge the Society of Exploration Geophysicists (SEG) and the SEG Advanced Modeling (SEAM) Corporation for creating the synthetic seismic dataset used in this work.

\subsection{Performance Comparisons}

Some of our performance improvements result from training and tuning the models on Google Cloud AI Platform. We utilized the TF-GAN library for all of our experiments. A summary of performance improvement results are shown in Table~\ref{tbl:PerformanceComparison}.

\begin{table}[h]
\centering
\caption{Performance Comparison on GPUs vs TPUs}
\label{tbl:PerformanceComparison}
\begin{tabular}{@{}llll@{}}
\toprule
 & Metric & GPU & TPU \\ \midrule
Performance & Evaluation Set - MSE & 0.022 & 0.00494 \\
Batch Size & Batch Size & 1 & 16 \\
Processing HW & Processing Hardware & 7  Tesla  P100s & Basic TPU \\
Speed & Samples/s & 4 samples/sec & 11 samples/sec \\
Cost & ML Units per 1k samples & 1.9 & 0.7 \\ \bottomrule
\end{tabular}
\end{table}

Table~\ref{tbl:PerformanceComparison} summarizes our results on the box noise dataset with a model trained on deterministic information. This was run for our baseline model with 2 repeating residual blocks in the generator and 2 blocks in the discriminator. The conditional information was added in via concatenation.

Table~\ref{tbl:SpeedUp} details results for a batch prediction run for 5000 3D Cubes.

\begin{table}[h]
\centering
\caption{Speed up comparison using GPU's for Batch Prediction}
\label{tbl:SpeedUp}
\begin{tabular}{lll}
\hline
\multicolumn{1}{|l|}{CPU Time(minutes)} & GPU Time(minutes) & GPU Speedup \\ \hline
208 & 15 & 13.8X \\ \hline
\end{tabular}
\end{table}

Reproducibility: The variance in SSIM is 0.05 on evaluation as observed after multiple runs at the end of full training.
\subsection{Additional information}

Table~\ref{tbl:HPTuning2} elaborates more on the information on hyperparameters tuned. We list a more specific range of parameters used for both 2D and 3D datasets. 

\begin{table}[hbt!]\label{hyperparameters_table}
    \centering
    \caption{Specific range of values used for hyperparameter tuning in 2D and 3D}
    \label{tbl:HPTuning2}
    \begin{tabular}{l|l|l|l|l}
    \hline
    Parameter & Min Value & Max Val & Scale \\ \hline
    2D Generator Depth & 16 & 40 & Linear \\ \hline
    2D Discriminator Depth & 6 & 18 & Linear \\ \hline
    2D Batch Size & 6 & 12 & Integer \\ \hline
    3D Generator Depth & 1 & 7 & Linear \\ \hline
    3D Discriminator Depth & 1 & 6 & Linear \\ \hline
    3D Batch Size & 1 & 4 & Integer \\ \hline
    \end{tabular}
\end{table}

\begin{figure}[ht]
  \centering
  \captionsetup{justification=centering}
  \begin{subfigure}[t]{0.18\linewidth}
    \includegraphics[width=\linewidth, height=5.5cm,valign=t]{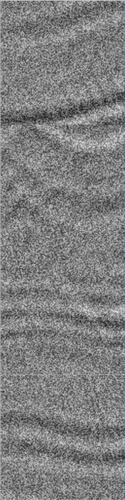}
    \caption{Degraded input.}
    \label{fig:LowRes}
  \end{subfigure}
  \begin{subfigure}[t]{0.18\linewidth}
    \includegraphics[width=\linewidth, height=5.5cm,valign=t]{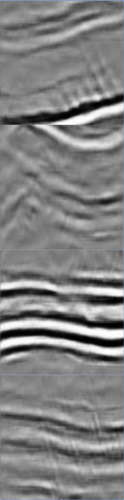}
    \caption{Baseline SRGAN.}
    \label{fig:VanillaSRGAN}
  \end{subfigure}
  \begin{subfigure}[t]{0.18\linewidth}
    \includegraphics[width=\linewidth, height=5.5cm,valign=t]{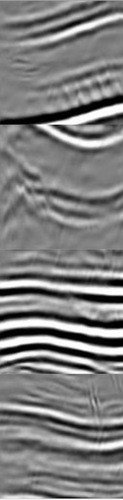}
    \caption{SRGAN with probabilistic information.}
    \label{fig:ProbSRGAN}
  \end{subfigure}
  \begin{subfigure}[t]{0.18\linewidth}
     \includegraphics[width=\linewidth, height=5.5cm,valign=t]{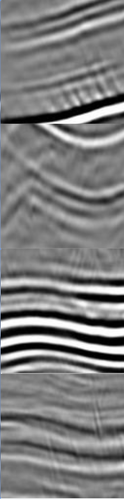}
    \caption{SRGAN with deterministic information.}
    \label{fig:DetSRGAN}
  \end{subfigure}
  \begin{subfigure}[t]{0.18\linewidth}
    \includegraphics[width=\linewidth, height=5.5cm,valign=t]{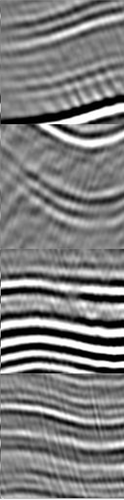}
    \caption{Ground truth.}
    \label{fig:GroundTruthSlices}
  \end{subfigure}
  \caption{Sample output of 2D models: Baseline SRGAN~\subref{fig:VanillaSRGAN}, probabilistic~\subref{fig:ProbSRGAN} and deterministic~\subref{fig:DetSRGAN} conditional SRGANs compared to the degraded input data~\subref{fig:LowRes} and ground truth~\subref{fig:GroundTruthSlices}. Both conditional SRGAN outputs show a marked decrease in coherent noise artifacts (diagonal vertical striping) compared to the ground truth image.}
  \label{fig:2DResults}
\end{figure}

Figure~\ref{fig:2DResults} offers a visual comparison of outputs of the 2D models we trained with the low-resolution input images and ground truth.

\subsection{Experiment setup}

We used Google Cloud AI Platform to run model training and hyperparameter tuning in the cloud, enabling us to leverage Tesla P100 GPUs and Google TPUs for training and Tesla P100 GPUs for inference. We used the TensorFlow GAN Estimator framework~\citep{tensorflow2015-whitepaper} to implement the model.

The models were trained for 100 000 steps with a batch size (hyperparameter) of 8 examples.
We used AI Platform Hyperparameter Tuning, which is based on Google Vizier~\citep{46180}, to optimize model performance.

\subsection{Open Source Implementation}

An open source implementation of the work of \citep{DBLP:journals/corr/LedigTHCATTWS16} is publicly available~\citep{tensorlayer2017}. This provides a similar base implementation of what we used for our Baseline SRGAN model.

\end{document}